\documentclass[11pt]{article}
\usepackage{blindtext}
\usepackage{graphicx}
\usepackage{amsmath}
\usepackage{amssymb}
\usepackage{amsthm}
\usepackage{xkeyval}
\title{The Commutation Relation for Cavity Mode Operators}
\author{Fesseha Kassahun\footnote{Email address: fesseha.kassahun@aau.edu.et.} \\
                Department of Physics,
                Addis Ababa University\\
                P. O. Box 33761, Addis Ababa, Ethiopia}

\begin{document}
\maketitle
\hspace*{20mm}We have obtained the commutation relation for a cavity~ mode driven by\newline
\hspace*{20mm}coherent light and interacting with a two-level atom. We have found that\newline
\hspace*{20mm}the commutation relation for a free cavity mode with or without photons\newline
\hspace*{20mm}is just the usual commutation~ relation. However, the commutation relat-\newline
\hspace*{20mm}ion for~ a cavity mode, which is interacting~ with a two-level~ atom, turns\newline
\hspace*{20mm}out to be different from ~the usual one. With ~the operator $\hat{c}$ defined by
$\hat{c}\newline
\hspace*{20mm}=\hat{a}+i\hat{b}$, the commutation relation for $\hat{c}$ and $\hat{c}^{\dagger}$ has
been also discussed.

\vspace*{10mm}
It appears that the usual commutation relation $[\hat{a},\hat{a}^{\dag}]=1$ [1] is taken to be the commutation relation for a cavity mode regardless of whether the cavity mode is interacting with an atom or not. In order to check the validity of this viewpoint, we seek here to obtain the commutation relation for a cavity mode driven by coherent light and interacting with a two-level atom. The interaction between the cavity mode and the driving coherent light can be described by the Hamiltonian
\begin{equation}\label{71}
\hat{H}'=i\lambda(\hat{b}\hat{a}^{\dagger}-\hat{b}^{\dagger}\hat{a}),
\end{equation}
where $\hat{a}$ the annihilation operator for the cavity mode, $\hat{b}$ is the annihilation operator for the driving coherent light, and $\lambda$ is the coupling constant between the cavity mode and the driving coherent light. In addition, the interaction between the cavity mode and the two-level atom can be described
at resonance by the Hamiltonian [2]
\begin{equation}\label{72}
\hat{H}^{\prime\prime}=ig(\hat{\sigma}^{\dagger}\hat{a}-\hat{a}^{\dag}\hat{\sigma}),
\end{equation}
where
\begin{equation}\label{73}
\hat{\sigma}=|b\rangle\langle a|
\end{equation}
is a lowering atomic operator and $g$ is the coupling constant between the cavity mode and the atom. Here $|a\rangle$ and $|b\rangle$ are the upper and lower levels of the two-level atom. We thus see that the Hamiltonian describing the interaction of the cavity mode with the driving coherent light and the two-level atom has the form
\begin{equation}\label{74}
\hat{H}=i\lambda(\hat{b}\hat{a}^{\dagger}-\hat{b}^{\dagger}\hat{a})+ig(\hat{\sigma}^{\dagger}\hat{a}-\hat{a}^{\dag}\hat{\sigma}).
\end{equation}
We consider the case in which the cavity mode is coupled to a vacuum reservoir via a single-port mirror. In addition, we carry out our calculation by taking into account the noise operators associated with the vacuum reservoir.

We recall that the operators in the Hamiltonian describing the interaction of a two-level atom with a single-mode light are fixed at the initial time. It then appears to be appropriate to fix the Hamiltonian given by (\ref{74}) at the initial time, whether we are working in the Schr$\ddot{o}$dinger or Heisenberg picture. The time evolution of the operator $\hat{a}(t)$ is described by the quantum Langevin equation [3]
\begin{equation}\label{75}
{d\over dt}\hat{a}(t) =-{\kappa\over 2}\hat{a}(t)-i[\hat{a}(t),\hat{H}(0)]+\hat{F}(t),
\end{equation}
where $\kappa$ is the cavity damping constant and $\hat{F}(t)$ is a noise operator with vanishing mean. In order to evaluate the commutator of $\hat{a}(t)$ and $\hat{H}(0)$, we introduce $\hat{\rho}(0)$, which is the density operator for the system under consideration at the initial time. We observe that $\hat{\rho}(0)[\hat{a}(t),\hat{H}(0)]$ is in the Heisenberg picture. This can also be written in the Schr$\ddot{o}$dinger picture as
\begin{equation}\label{76}
\hat{\rho}(0)[\hat{a}(t),\hat{H}(0)]=\hat{\rho}(t)[\hat{a}(0),\hat{H}(0)].
\end{equation}
Now applying the Hamiltonian given by (\ref{74}), we find
\begin{equation}\label{77}
[\hat{a}(t),\hat{H}(0)]=i\lambda\hat{b}(t)-ig\hat{\sigma}(t).
\end{equation}
Hence in view of this result, Eq. (\ref{75}) takes the form
\begin{equation}\label{78}
{d\over dt}\hat{a}(t)=-{\kappa\over 2}\hat{a}(t)+\lambda\hat{b}(t)-g\hat{\sigma}(t)+\hat{F}(t).
\end{equation}
With the intention of obtaining the solution of this equation, we replace $\hat{b}(t)$ by the c-number $\beta$, considered to be real and constant. Thus making use of this replacement, Eq. (\ref{78}) can be put in the form
\begin{equation}\label{79}
{d\over dt}\hat{a}(t)=-{\kappa\over 2}\hat{a}(t)-g\hat{\sigma}(t)+\varepsilon+\hat{F}(t),
\end{equation}
in which $\varepsilon=\lambda\beta$.

In addition, employing the relation
\begin{equation}\label{710}
{d\over dt}\langle\hat{A}(t)\rangle=-i\langle[\hat{A}(t),\hat{H}(0)]\rangle
\end{equation}
along with Eq. (\ref{74}), one readily obtains (with the time argument suppressed)
\begin{equation}\label{711}
{d\over dt}\langle\hat{\sigma}\rangle=g\langle(\hat{\eta}_{b}-\hat{\eta}_{a})a\rangle,
\end{equation}
\begin{equation}\label{712}
{d\over dt}\langle\hat{\eta}_{a}\rangle=g\langle\hat{\sigma}^{\dagger}\hat{a}\rangle +g\langle\hat{a}^{\dag}\hat{\sigma}\rangle,
\end{equation}
where
\begin{equation}\label{713}
\hat{\eta}_{a}=|a\rangle\langle a|,
\end{equation}
\begin{equation}\label{714}
\hat{\eta}_{b}=|b\rangle\langle b|.
\end{equation}

We see that Eqs. (\ref{711}) and (\ref{712}) are nonlinear differential equations and hence it is not possible to obtain exact solutions of these equations. We seek to overcome this problem by making use of the so-called large-time approximation scheme [4]. Hence applying this approximation scheme, we obtain from Eq. (\ref{79}) the approximately valid relation
\begin{equation}\label{715}                        \hat{a}(t)=-{2g\over\kappa}\hat{\sigma}(t)+{2\varepsilon\over\kappa}+{2\over\kappa}\hat{F}(t).
\end{equation}
Now substitution of (\ref{715}) into the aforementioned equations yields
\begin{equation}\label{716}
{d\over dt}\langle\hat{\sigma}\rangle=-{1\over 2}\gamma_{c}\langle\hat{\sigma}\rangle+{2g\varepsilon\over\kappa}\langle\hat{\eta}_{b}-\hat{\eta}_{a}\rangle
+{2g\over\kappa}\langle(\hat{\eta}_{b}-\hat{\eta}_{a})\hat{F}\rangle,
\end{equation}
\begin{equation}\label{717}
{d\over dt}\langle\hat{\eta}_{a}\rangle=-\gamma_{c}\langle\hat{\eta}_{a}\rangle+{2g\varepsilon\over\kappa}\langle\hat{\sigma}
+\hat{\sigma}^{\dagger}\rangle+{2g\over\kappa}\langle\hat{\sigma}^{\dag}\hat{F}+\hat{F}^{\dag}\hat{\sigma}\rangle,
\end{equation}
where
\begin{equation}\label{718}
\gamma_{c}=4g^{2}/\kappa.
\end{equation}
Assuming that the atomic and noise operators are not correlated, we have
\begin{equation}\label{719}
\langle(\hat{\eta}_{b}-\hat{\eta}_{a})\hat{F}\rangle=\langle\hat{\eta}_{b}-\hat{\eta}_{a}\rangle\langle\hat{F}\rangle=0,
\end{equation}
\begin{equation}\label{720}
\langle\hat{\sigma}^{\dag}\hat{F}\rangle=\langle\hat{\sigma}^{\dag}\rangle\langle\hat{F}\rangle=0,
\end{equation}
\begin{equation}\label{721}
\langle\hat{F}^{\dag}\hat{\sigma}\rangle=\langle\hat{F}^{\dag}\rangle\langle\hat{\sigma}\rangle=0.
\end{equation}
We therefore see that
\begin{equation}\label{722}
{d\over dt}\langle\hat{\sigma}\rangle=-{1\over 2}\gamma_{c}\langle\hat{\sigma}\rangle+{2g\varepsilon\over\kappa}\langle\hat{\eta}_{b}-\hat{\eta}_{a}\rangle.
\end{equation}
\begin{equation}\label{723}
{d\over dt}\langle\hat{\eta}_{a}\rangle=-\gamma_{c}\langle\hat{\eta}_{a}\rangle+{2g\varepsilon\over\kappa}\langle\hat{\sigma}
+\hat{\sigma}^{\dagger}\rangle.
\end{equation}
Moreover, we note that the steady-state solutions of Eqs. (\ref{722}) and (\ref{723}) have the form
\begin{equation}\label{724}
\langle\hat{\sigma}\rangle={4g\varepsilon\over\kappa\gamma_{c}}\langle\hat{\eta}_{b}-\hat{\eta}_{a}\rangle,
\end{equation}
\begin{equation}\label{725}
\langle\hat{\eta}_{a}\rangle={2g\varepsilon\over\kappa\gamma_{c}}\langle\hat{\sigma}+\hat{\sigma}^{\dagger}\rangle.
\end{equation}
Hence on substituting (\ref{724}) into Eq. (\ref{725}), we find
\begin{equation}\label{726}
\langle\hat{\eta}_{a}\rangle={4\varepsilon^{2}\over\kappa\gamma_{c}}\langle\hat{\eta}_{b}-\hat{\eta}_{a}\rangle.
\end{equation}
Now taking into account the completeness relation
\begin{equation}\label{727}
\hat{\eta}_{a}+\hat{\eta}_{b}=\hat{I},
\end{equation}
we get
\begin{equation}\label{728}
\langle\hat{\eta}_{a}\rangle={4\varepsilon^{2}\over{8\varepsilon^{2}+\kappa\gamma_{c}}}
\end{equation}
and applying once more Eq. (\ref{727}), we have
\begin{equation}\label{729}
\langle\hat{\eta}_{b}\rangle=1-{4\varepsilon^{2}\over{8\varepsilon^{2}+\kappa\gamma_{c}}}.
\end{equation}
It is not hard to realize that $\langle\hat{\eta}_{a}\rangle$ and $\langle\hat{\eta}_{a}\rangle$ represent
the probabilities for the two-level atom to be in the upper and lower levels, respectively.
Finally, with the aid of (\ref{728}) and (\ref{729}), one can put Eq. (\ref{724}) in the form
\begin{equation}\label{730}
\langle\hat{\sigma}\rangle={4g\varepsilon\over{8\varepsilon^{2}+\kappa\gamma_{c}}}.
\end{equation}

Furthermore, applying the relation
\begin{equation}\label{731}
{d\over dt}\langle\hat{a}(t)\hat{a}^{\dagger}(t)\rangle=\bigg\langle{d\hat{a}(t)\over dt}\hat{a}^{\dagger}(t)\bigg\rangle+
\bigg\langle\hat{a}(t){d\hat{a}^{\dagger}(t)\over dt}\bigg\rangle
\end{equation}
along with Eq. (\ref{79}) and its adjoint, we readily get
\begin{eqnarray}\label{732}
{d\over dt}\langle\hat{a}(t)\hat{a}^{\dagger}(t)\rangle\hspace*{-3mm}&=\hspace*{-3mm}&-\kappa\langle\hat{a}(t)\hat{a}^{\dagger}(t)\rangle
+\varepsilon\langle\hat{a}(t)+\hat{a}^{\dagger}(t)\rangle-g\big[\langle\hat{\sigma}(t)\hat{a}^{\dagger}(t)\rangle+\langle\hat{a}(t)\hat{\sigma}^{\dagger}(t)\rangle
\nonumber\\&&
+\langle\hat{F}(t)\hat{a}^{\dagger}(t)\rangle+\langle\hat{a}(t)\hat{F}^{\dagger}(t)\rangle\big].
\end{eqnarray}
It proves to be convenient to replace the operators $\hat{a}^{\dagger}(t)$ and $\hat{a}(t)$ that appear in the second, third, fourth, and fifth terms in Eq. (\ref{732}) by expression (\ref{715}) and its adjoint. We then find
\begin{eqnarray}\label{733}
{d\over dt}\langle\hat{a}(t)\hat{a}^{\dagger}(t)\rangle\hspace*{-3mm}&=\hspace*{-3mm}&-\kappa\langle\hat{a}(t)\hat{a}^{\dagger}(t)\rangle
+\gamma_{c}\langle\hat{\eta}_{b}\rangle+{4\varepsilon^{2}\over\kappa}-{4g\varepsilon\over\kappa}\langle\hat{\sigma}(t) +\hat{\sigma}^{\dagger}(t)\rangle\nonumber\\&&
-{2g\over\kappa}\big(\langle\hat{\sigma}(t)\hat{F}^{\dagger}(t)\rangle
+\langle\hat{F}(t)\hat{\sigma}^{\dagger}(t)\rangle\big)
+\langle\hat{F}(t)\hat{a}^{\dagger}(t)\rangle
+\langle\hat{a}(t)\hat{F}^{\dagger}(t)\rangle,
\end{eqnarray}
so that in view of the assumption that the atomic and noise operators are not correlated, there follows
\begin{eqnarray}\label{734}
{d\over dt}\langle\hat{a}(t)\hat{a}^{\dagger}(t)\rangle\hspace*{-3mm}&=\hspace*{-3mm}&-\kappa\langle\hat{a}(t)\hat{a}^{\dagger}(t)\rangle
+\gamma_{c}\langle\hat{\eta}_{b}\rangle+{4\varepsilon^{2}\over\kappa}-{4g\varepsilon\over\kappa}\langle\hat{\sigma}(t) +\hat{\sigma}^{\dagger}(t)\rangle
\nonumber\\&&
+\langle\hat{F}(t)\hat{a}^{\dagger}(t)\rangle
+\langle\hat{a}(t)\hat{F}^{\dagger}(t)\rangle.
\end{eqnarray}
Moreover, the solution of Eq. (\ref{79}) can be written as
\begin{equation}\label{735}
\hat{a}(t)=\hat{a}(0)e^{-\kappa t/2}+e^{-\kappa t/2}\int^{t}_{0}e^{\kappa t'/2}[-g\hat{\sigma}(t')+\varepsilon+\hat{F}(t')]dt',
\end{equation}
and multiplying this equation on the right by $\hat{F}^{\dagger}(t)$, we have
\begin{equation}\label{736}
\langle\hat{a}(t)\hat{F}^{\dagger}(t)\rangle=\langle\hat{a}(0)\hat{F}^{\dagger}(t)\rangle e^{-\kappa t/2}+e^{-\kappa t/2}\int^{t}_{0}e^{\kappa t'/2}[-g\langle\hat{\sigma}(t')\hat{F}^{\dagger}(t)\rangle
+\langle\hat{F}(t')\hat{F}^{\dagger}(t)\rangle]dt',
\end{equation}
in which we have used the fact that $\varepsilon\langle\hat{F}^{\dagger}(t)\rangle=0$.
Now in view of the assumptions
\begin{equation}\label{737}
\langle\hat{a}(0)\hat{F}^{\dagger}(t)\rangle=0
\end{equation}
and
\begin{equation}\label{738}
\langle\hat{\sigma}(t')\hat{F}^{\dagger}(t)\rangle=0,
\end{equation}
Eq. (\ref{736}) takes the form
\begin{equation}\label{739}
\langle\hat{a}(t)\hat{F}^{\dagger}(t)\rangle=e^{-\kappa t/2}\int^{t}_{0}e^{\kappa t'/2}\langle\hat{F}(t')\hat{F}^{\dagger}(t)\rangle dt'.
\end{equation}
Hence using the correlation property [5]
\begin{equation}\label{740}
\langle\hat{F}(t')\hat{F}^{\dagger}(t)\rangle=\kappa\delta(t-t'),
\end{equation}
we arrive at
\begin{equation}\label{741}
\langle\hat{a}(t)\hat{F}^{\dagger}(t)\rangle={\kappa\over 2}.
\end{equation}
Therefore, on substituting (\ref{741}) and its complex conjugate into Eq. (\ref{734}), there follows
\begin{equation}\label{742}
{d\over dt}\langle\hat{a}(t)\hat{a}^{\dagger}(t)\rangle=-\kappa\langle\hat{a}(t)\hat{a}^{\dagger}(t)\rangle
+\gamma_{c}\langle\hat{\eta}_{b}\rangle+{4\varepsilon^{2}\over\kappa}-{4g\varepsilon\over\kappa}\langle\hat{\sigma}(t) +\hat{\sigma}^{\dagger}(t)\rangle+\kappa.
\end{equation}

The steady-state solution of Eq. (\ref{742}) has the form
\begin{equation}\label{743}
\langle\hat{a}\hat{a}^{\dagger}\rangle={\gamma_{c}\over\kappa}\langle\hat{\eta}_{b}\rangle+{4\varepsilon^{2}\over\kappa^{2}}
-{4g\varepsilon\over\kappa^{2}}\langle\hat{\sigma}+\hat{\sigma}^{\dagger}\rangle+1
\end{equation}
and on taking into account (\ref{730}), we obtain
\begin{equation}\label{744}
\langle\hat{a}\hat{a}^{\dagger}\rangle={\gamma_{c}\over\kappa}\langle\hat{\eta}_{b}\rangle+{4\varepsilon^{2}\over\kappa^{2}}
-{\gamma_{c}\over\kappa}{8\varepsilon^{2}\over{8\varepsilon^{2}+\kappa\gamma_{c}}}+1.
\end{equation}
Following the same procedure, we can also readily establish that at steady state
\begin{equation}\label{745}
\langle\hat{a}^{\dagger}\hat{a}\rangle={\gamma_{c}\over\kappa}\langle\hat{\eta}_{a}\rangle+{4\varepsilon^{2}\over\kappa^{2}}
-{\gamma_{c}\over\kappa}{8\varepsilon^{2}\over{8\varepsilon^{2}+\kappa\gamma_{c}}}.
\end{equation}
Now employing Eqs. (\ref{744}) and (\ref{745}), we easily find
\begin{equation}\label{746}
[\hat{a},\hat{a}^{\dagger}]={\gamma_{c}\over\kappa}(\langle\hat{\eta}_{b}\rangle-\langle\hat{\eta}_{a}\rangle)+1.
\end{equation}
Finally, with the aid of Eqs. (\ref{728}) and (\ref{729}), one can put Eq. (\ref{746}) in the form
\begin{equation}\label{747}
[\hat{a},\hat{a}^{\dagger}]={\gamma_{c}^{2}\over{8\varepsilon^{2}+\kappa\gamma_{c}}}+1
\end{equation}
and on account of (\ref{728}) the mean photon number of the cavity mode has the form
\begin{equation}\label{748}
\overline{n}={4\varepsilon^{2}\over\kappa^{2}}
-{\gamma_{c}\over\kappa}{4\varepsilon^{2}\over{8\varepsilon^{2}+\kappa\gamma_{c}}}.
\end{equation}

We realize that Eq. (\ref{747}) represents the commutation relation for a cavity mode which is interacting with a two-level atom. And we notice that the first term on the right side of this equation is due to the interaction of the cavity mode with the two-level atom and the second term is due to the vacuum reservoir noise.
We now consider the case in which the cavity mode is not interacting with an atom. Then for this case (g=0), Eq. (\ref{747}) takes the form
\begin{equation}\label{749}
[\hat{a},\hat{a}^{\dagger}]=1,
\end{equation}
with
\begin{equation}\label{750}
\overline{n}={4\varepsilon^{2}\over\kappa^{2}}.
\end{equation}
This represents the commutation relation for a free cavity mode with photons.
Alternatively, in the absence of the driving coherent light ($\varepsilon=0$), Eq. (\ref{747}) has the form
\begin{equation}\label{751}
[\hat{a},\hat{a}^{\dagger}]={\gamma_{c}\over\kappa}+1,
\end{equation}
with $\overline{n}=0$. We see from Eq. (\ref{729}) that the atom in this case is in the lower level. We observe that Eq. (\ref{751}) represents the commutation relation for a vacuum cavity mode which is interacting with a two-level atom. We may envisage this interaction as a process in which virtual photons are absorbed and emitted, with the atom continuing to be in the lower level. We note that in the absence of this interaction (g=0), Eq. (\ref{751}) goes over into
\begin{equation}\label{752}
[\hat{a},\hat{a}^{\dagger}]=1.
\end{equation}
We now realize that the usual commutation relation is just the commutation relation for a free cavity mode with or without photons.

Furthermore, we wish to discuss the commutation relation for the superposition of two light modes, represented by the operators $\hat{a}$ and $\hat{b}$. It proves to be very convenient to represent the superposed light modes by the operator $\hat{c}$ defined by
\begin{equation}\label{753}
\hat{c}=\hat{a}+i\hat{b}.
\end{equation}
We thus see that
\begin{equation}\label{754}
[\hat{c},\hat{c}^{\dagger}]=[\hat{a},\hat{a}^{\dagger}]+[\hat{b},\hat{b}^{\dagger}]+i[\hat{b},\hat{a}^{\dagger}]
-i[\hat{a},\hat{b}^{\dagger}].
\end{equation}
We assume that the operators $\hat{a}$ and $\hat{b}$ commute. It then follows that
\begin{equation}\label{755}
[\hat{c},\hat{c}^{\dagger}]=[\hat{a},\hat{a}^{\dagger}]+[\hat{b},\hat{b}^{\dagger}].
\end{equation}
We therefore note that the commutator for the superposed light modes is just the sum of the commutators for the separate light modes.

Finally, with the intention to determine the conditions under which the definition given by (\ref{753}) holds, we consider the mean photon number of the superposed light modes. Applying Eq. (\ref{753}), this mean photon number can be written as
\begin{equation}\label{756}
\langle\hat{c}^{\dagger}\hat{c}\rangle=\langle\hat{a}^{\dagger}\hat{a}\rangle+\langle\hat{b}^{\dagger}\hat{b}\rangle
+i(\langle\hat{a}^{\dagger}\hat{b}\rangle-\langle\hat{b}^{\dagger}\hat{a}\rangle).
\end{equation}
We expect the mean photon number to be the sum of the mean photon numbers of the separate light modes. We then demand that the last term in Eq. (\ref{756}) must be zero. To achieve this, we first assert that the superposed light modes must not be correlated. In view of this, the mean photon number can be rewritten as
\begin{equation}\label{757}
\langle\hat{c}^{\dagger}\hat{c}\rangle=\langle\hat{a}^{\dagger}\hat{a}\rangle+\langle\hat{b}^{\dagger}\hat{b}\rangle
+i(\langle\hat{a}^{\dagger}\rangle\langle\hat{b}\rangle-\langle\hat{b}^{\dagger}\rangle\langle\hat{a}\rangle).
\end{equation}
We now realize that the last term would vanish if
$\langle\hat{a}\rangle=\langle\hat{b}\rangle=0$ or $\langle\hat{a}\rangle=\varepsilon_{a}$ and $\langle\hat{b}\rangle=\varepsilon_{b}$, with $\varepsilon_{a}$ and $\varepsilon_{b}$ taken to be real or equal.

\begin{picture}(200,50)
\put (200,10){\line(1,0){80}}
\end{picture}
\vspace*{3mm}

\noindent
[1] P. L. Knight and B. W. Shore, Phys. Rev. \textbf{A} 48, 642 (1993). \newline
[2] Fesseha Kassahun, Quantum Analysis of Light, (Kindle Direct Publishing, 2018).\newline
[3] M. J. Collett and D. F. Walls, Phys. Rev. \textbf{A} 32, 2887 (1985).\newline
[4] Fesseha Kassahun, Opt. Commun. 284, 1357 (2011).\newline
[5] C. W. Gardiner and M. J. Collett, Phys. Rev. \textbf{A} 31, 3761 (1985).

\end{document}